\documentclass[twocolumn,prb,superscriptaddress,showpacs,showkeys,amsmath,amssymb,floatfix,aps]{revtex4}
\usepackage[english]{babel}
\usepackage[latin1]{inputenc}
\usepackage{graphics,graphicx,epsfig}
\usepackage{dcolumn}

\newcommand {\eu} {EuNiO$_3$}
\newcommand {\pr} {PrNiO$_3$}
\newcommand {\nd} {NdNiO$_3$}
\newcommand {\rn} {$R$NiO$_3$}
\newcommand {\yn} {YNiO$_3$}
\newcommand {\lu} {LuNiO$_3$}

\newcommand {\n} {Ni$^{3+}$}
\newcommand {\tmi} {T$_{MI}$}

\newcommand {\ld} {L$_{2}$}
\newcommand {\lt} {L$_{3}$}

\newcommand {\so} {$l\cdot s$}

\begin{document}


\title{Spin-orbit induced mixed-spin ground state in $R$NiO$_3$ perovskites probed by XAS: new insight into the metal to insulator transition}

\author{C. Piamonteze}
\affiliation{Laboratório Nacional Luz Síncrotron, Caixa Postal
6192, 13084-971, Campinas/SP, Brazil} \affiliation{IFGW/UNICAMP,
13083-970, Campinas/SP, Brazil}

\author{F. M. F. de Groot}
\affiliation{Department of Inorganic Chemistry and Catalysis,
Utrecht University, Sorbonnelaan 16, 3584 CA, Utrecht, The
Netherlands}

\author{H. C. N. Tolentino}
\affiliation{Laboratório Nacional Luz Síncrotron, Caixa Postal
6192, 13084-971, Campinas/SP, Brazil}

\author{A. Y. Ramos}
\affiliation{Laboratório Nacional Luz Síncrotron, Caixa Postal
6192, 13084-971, Campinas/SP, Brazil} \affiliation{LMCP - CNRS,
Université de Paris 6, Paris, France}

\author{N.  E.  Massa}
\affiliation{Laboratorio Nacional de Investigación y Servicios en
Espectroscopía Optica, Centro CEQUINOR, UNLP, 1900 La Plata,
Argentina}

\author{J.  A.  Alonso}
\author{M.  J. Mart\'{\i}nez-Lope}
\affiliation{Instituto de Ciencia de Materiales de Madrid,
C.S.I.C, Cantoblanco, E-28049 Madrid, Spain}

\date{\today}

\begin{abstract}
We report on a Ni L$_{2,3}$ edges x-ray absorption spectroscopy
(XAS) study in $R$NiO$_3$ perovskites. These compounds exhibit a
metal to insulator ($MI$) transition as temperature decreases. The
L$_{3}$ edge presents a clear splitting in the insulating state,
associated to a less hybridized ground state. Using charge
transfer multiplet calculations, we establish the importance of
the crystal field and 3d spin-orbit coupling to create a
mixed-spin ground state. We explain the $MI$ transition in
$R$NiO$_3$ perovskites in terms of modifications in the Ni$^{3+}$
crystal field splitting that induces a spin transition from an
essentially low-spin (LS) to a mixed-spin state.
\end{abstract}



\maketitle

Rare-earth nickel perovskites (\rn, $R$=rare earth) present a
sharp well-defined metal to insulator ($MI$) transition as
temperature decreases \cite{Lacorre1991}. The transition
temperature, \tmi, increases with reducing the $R$ ion size, which
determines the degree of distortion of the structure
\cite{Munoz1992}. It was proposed that the gap opening would be
due to a smaller Ni-O-Ni superexchange angle leading to a
reduction of the bandwidth \cite{Torrance1992}. However,
non-negligible electron-phonon interactions \cite{Massa1997b} and
a shift in \tmi\ with oxygen isotope substitution
\cite{Medarde1998} evidenced the importance of modifications in
Ni-O interatomic distances, suggesting a phonon assisted mechanism
for conduction. As temperature decreases, these nickelates undergo
a magnetic transition to an unusual antiferromagnetic order
\cite{Munoz1994,Carvajal1998,Alonso1999}. The magnetic arrangement
for the lighter $R$ compounds ($R$=Pr, Nd, Sm, Eu) was refined
with a single Ni moment (0.9$\mu_B$) and required non-equivalent
couplings among Ni ions to stabilize the structure
\cite{Munoz1994,Carvajal1998}. This is a quite unusual situation
in an orthorhombic crystallographic structure whose Ni sites are
all equivalent\cite{Munoz1992}. For the heavier $R$ compounds,
Alonso {\it et al.} \cite{Alonso1999,Alonso2000} established a
monoclinic distortion in the crystallographic structure leading to
two different Ni sites with longer and shorter Ni-O distances
alternating along the three axis. The antiferromagnetic structure
was explained by a charge ordering defined among the different Ni
sites, each one with different magnetic moments (1.4 and
0.7$\mu_B$ for YNiO$_3$) \cite{Alonso1999}. More recently, some
evidences that the low temperature distortion is shared by all
members of the \rn\ family were reported \cite{delaCruz2002}.
Concerning the Ni local structure, our recent EXAFS results
demonstrated the existence of two different Ni sites in all \rn\
compounds, regardless their long-range crystallographic structure
\cite{PiamontezePhyScri,PiamontezeLetterEXAFS}.

From spectroscopic data together with configuration interaction
calculations, Mizokawa {\it et al.} \cite{Mizokawa1995}
established for \pr\ at room temperature a metallic ground state
composed of 34\%3d$^7$+56\%3d$^8$\underline{\bf
L}+10\%3d$^9$\underline{\bf L}$^2$, where \underline{\bf L} stands
for a ligand hole. Owing to the high degree of hybridization in
the ground state, with hole-transfer from Ni3d to O2p orbitals,
these compounds have been classified as self-doped Mott insulators
\cite{Mizokawa2000}. Such a mixed metallic ground state, mostly
3d$^8$\underline{\bf L}, is compatible with a \n\ low-spin
configuration, because the amount of charge transfer for parallel
spin almost equals that for antiparallel spin \cite{Mizokawa1995}.
However, the spectral shape in \rn\ compounds is very sensitive to
the transition from metallic to insulating states
\cite{PiamontezeSRL} and a complete description of the spin degree
of freedom remains to be given. As in recent outcomes on Co$^{3+}$
oxides \cite{Hu2004,Nekrasov2003}, where unconventional spin
states exist due to the competition between crystal field
splitting and effective 3d exchange interaction, assignments made
so far about \n\ in \rn\ compounds have to be reexamined.

We report here Ni L-edge absorption measurements, which probes
directly the available Ni3d states, together with charge transfer
multiplet calculations. We establish the importance of the crystal
field and 3d spin-orbit (\so) coupling to create a mixed-spin
ground state. We explain the $MI$ transition in \rn\ perovskites
in terms of modifications in the \n\ crystal field splitting that
localizes the electronic states and induces a spin transition from
an essentially low-spin (LS) to a mixed-spin state, with a quite
large contribution from a high-spin (HS) state.

\begin{figure}
\begin{center}
\epsfig{file=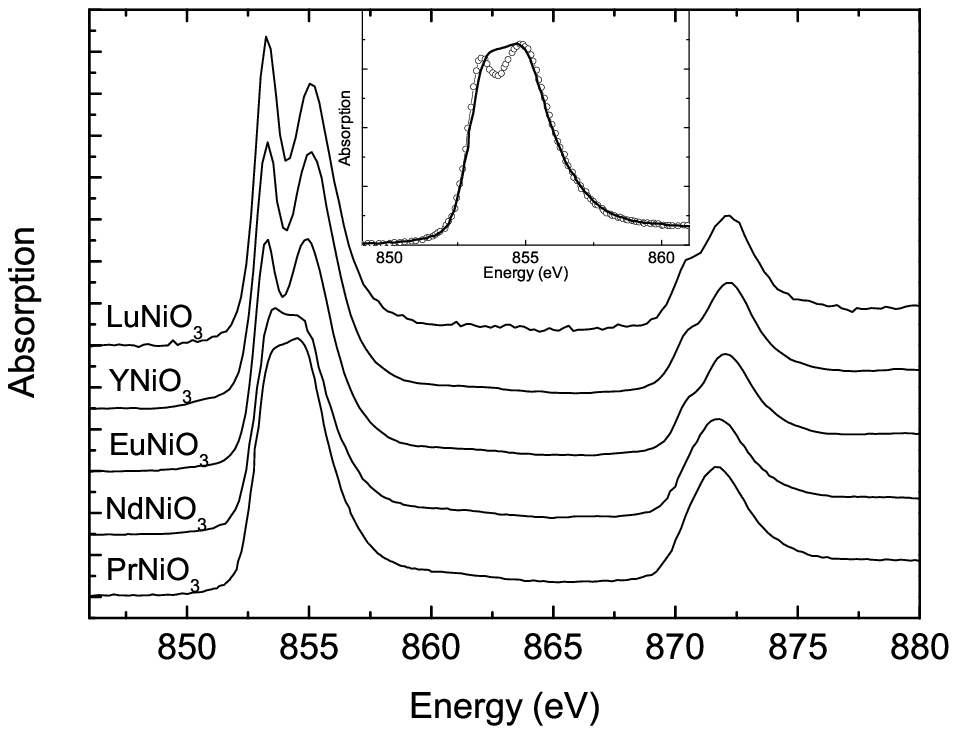,width=7.0cm}
\caption{\label{experimental}Ni L$_{2,3}$ edge spectra at 300K for
\pr, \nd, \eu, \yn\ and \lu, whose \tmi\ are 130, 200, 462, 582
and 599K, respectively. Inset: \lt\ edge spectra for \pr\ at 300K
(solid line) and at 96K (open circles).}
\end{center}
\end{figure}

Soft x-ray absorption measurements at the Ni L$_{2,3}$ edges were
carried out at the SGM beamline of LNLS, Brazil, using a spherical
grating monochromator with an energy resolution of 0.8eV at 845eV.
Data were collected at 300K and 96K using a liquid nitrogen cold
finger. The bulk polycrystalline \rn\ ($R$=Pr, Nd, Eu, Y and Lu)
samples were obtained by a wet-chemistry technique, as described
elsewhere \cite{Alonso1995b}. The experimental spectra measured at
300K for \lu, \yn, \eu, \nd\ and \pr\ are shown in figure
\ref{experimental}. The $MI$ transition for these compounds takes
place at 599K, 582K, 460K, 200K and 130K, respectively
\cite{Alonso2000}. At 300K, \nd\ and \pr\ are at the metallic
state whereas the other compounds are at the insulating state. For
the insulators, a clear splitting is observed at the \lt\ edge
($\sim$854eV), while for the metals this splitting is absent.
Differences among spectra measured in the different electronic
states are also noticeable at the \ld\ edge ($\sim$872eV). \pr\
and \nd\ compounds spectra measured at 96K, when both are
insulators, are analogous to those of $R$=Eu, Y and Lu at 300K, as
shown by the splitting at the \lt\ edge for the \pr\ compound
(inset fig.\ref{experimental}). Such modifications in the
experimental spectra, unambiguously associated to the $MI$
transition, demonstrate that important changes take place at the
Ni local electronic structure. As the spectral shape depends
mostly if the compound is metal or insulator, we conclude that the
short range scale electronic structure of all compounds shares a
common basis.

\begin{figure}
\begin{center}
\epsfig{file=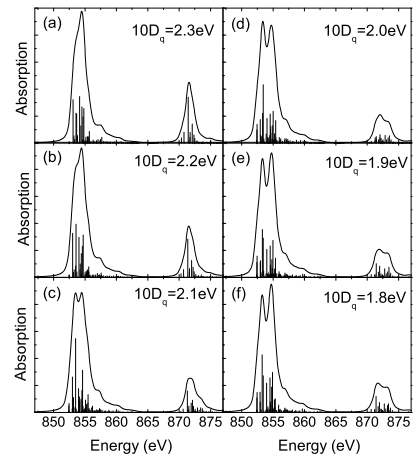,width=6.0cm}
\caption{\label{sim10dq} CTM calculations of the Ni L XAS spectra
in $D_{4h}$ symmetry. The value of $10D_q$ is varied between 2.3
eV (a) and 1.8 eV (f), in steps of 0.1 eV. The other parameters
used are described in the text.}
\end{center}
\end{figure}

To identify the electronic interactions accounting for the
experimental data, charge transfer (configuration interaction)
multiplet (CTM) \cite{deGroot1994,deGroot2004} calculations were
carried out. The calculations take into account interactions
between three configurations, 3d$^7$, 3d$^8$\underline{\bf L}, and
3d$^9$\underline{\bf L}$^2$, in $D_{4h}$ symmetry. A series of
calculations varying the crystal field splitting parameter
$10D_q$, keeping $D_s=0.1eV$, $D_t=0.2eV$ fixed, is presented in
figure \ref{sim10dq}. The configuration interaction parameters are
$\Delta=0.5eV$, $U_{dd}-U_{pd}=-1eV$ and the transfer integrals
$T(B_1)=T(A_1)=2T(B_2)=2T(E)=2eV$. The calculations matching the
best with the experimental spectra for the metallic compounds are
those with $10D_q=2.2eV$. The \n\ metallic ground state turns out
to be composed of 49\%3d$^7$+47\%3d$^8$\underline{\bf
L}+4\%3d$^9$\underline{\bf L}$^2$. The relative proportions for
the three configurations are in reasonable agreement with those
found by Mizokawa {\it et al.} for metallic \pr\
\cite{Mizokawa1995}. The slight difference can be partially
ascribed to the way the ligand hole is treated in the codes
\cite{comment}. For the insulating compounds, a very good
agreement with experiments is achieved by decreasing $10D_q$ down
to $2.0$ or $1.9eV$. Such a decrease in 10D$_q$ leads to a less
hybridized ground state composed of
61\%3d$^7$+37\%3d$^8$\underline{\bf L}+2\%3d$^9$\underline{\bf
L}$^2$.

\begin{figure}
\begin{center}
\epsfig{file=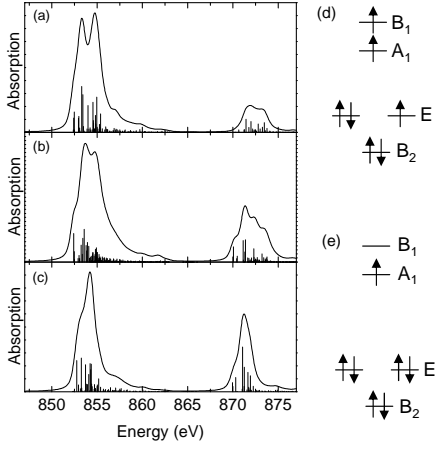,width=6.0cm}
\caption{\label{gsee}CTM calculations for $10D_q=1.9eV$ (a) with
\so\ coupling,(b) without \so\ coupling and (c) without \so\
coupling but with the initial state being the first excited state;
(d) Scheme of HS - $^4$E and (e) LS - $^2$A$_1$ states.}
\end{center}
\end{figure}

Figure \ref{gsee}a shows the simulated spectrum for $10D_q=1.9eV$
taking into account the \so\ coupling, as in fig.\ref{sim10dq}e,
while figure \ref{gsee}b shows the calculation without \so\
coupling. The effect on the spectral shape is remarkable. The
ground state without \so\ coupling is a HS state, schematically
represented in figure \ref{gsee}d. Figure \ref{gsee}c shows a
calculation without \so\ coupling, but when the initial state of
the transition is the first excited state, which is a LS state,
schematically represented in figure \ref{gsee}e. The energy
difference between these two states is found to be $\sim0.1eV$.
Since the \so\ coupling is of the same order, the insulating
ground state turns out to be a mixing of LS and HS states. In
addition, the calculation of the metallic state
(fig.\ref{sim10dq}b) is quite similar to that shown in
fig.\ref{gsee}c, indicating that the LS state gives the main
contribution to the metallic ground state. We conclude that the
ground state composition is very sensitive to the spin-orbit (\so)
coupling and that modifications in the crystal field splitting
(10D$_q$) change the relative contribution of both states to the
mixed ground state.

The ground state (GS) can be decomposed into different symmetries
that are mixed up by the \so\ coupling:

\begin{eqnarray}
\Psi_{GS}&=&\sum_{i; k+l+m+n=3}\alpha_i \left| a_1^k b_1^l b_2^m
e^n\right\rangle \nonumber\\&+& \sum_{j; k'+l'+m'+n'=2} \alpha_j
\left| a_1^{k'} b_1^{l'} b_2^{m'} e^{n'}\underline{\bf
L}\right\rangle \nonumber\\&+& \sum_{p; k''+l''+m''+n''=1}
\alpha_p \left| a_1^{k''} b_1^{l''} b_2^{m''}
e^{n''}\underline{\bf L}^2\right\rangle \label{eq_gs}
\end{eqnarray}

\begin{figure}
\begin{center}
\epsfig{file=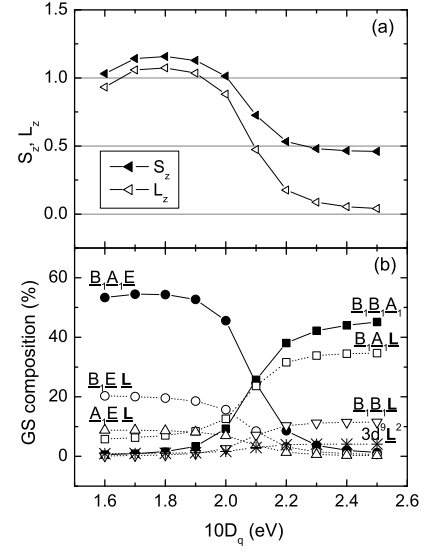,width=6.0cm}
\caption{\label{spings}(a) Variation of $S_z$, $L_z$ and (b)
ground state composition as a function of $10D_q$. The
configuration \underline{$B_1B_1A_1$} corresponds to the LS
$^2A_1$ state in fig.\ref{gsee}e and the configuration
\underline{$B_1A_1E$} corresponds to the HS $^4E$ state in
fig.\ref{gsee}d. The others correspond to charge transfer
configurations.}
\end{center}
\end{figure}

In this expression $k+l+m+n$, $k'+l'+m'+n'$ and $k''+l'+m''+n''$
are the number of d holes in 3d$^7$, 3d$^8$\underline{\bf L} and
3d$^9$\underline{\bf L}$^2$ configurations, respectively. The
$\alpha_i$, $\alpha_j$, $\alpha_p$ values can be obtained as
described by Wasinger {\it et al.} \cite{Wasinger2003}. The
contributions of these different symmetries to the ground state
and to spin ($S_z$) and orbital angular ($L_z$) moments as
function of $10D_q$ are shown in figure \ref{spings}. The labels
in figure \ref{spings}b refer to the distribution of the d holes
among $B_1$, $A_1$, $B_2$ and $E$ orbitals. At the metallic state,
with $10D_q=2.2eV$, the ground state is composed of $\sim$38\%\
\underline{$B_1B_1A_1$} (LS) and a small amount ($\sim$8\%) of
\underline{$B_1A_1E$} (HS), the rest being charge transfer
configurations. This leads to a net spin moment $S_z\sim0.5$, in
accordance with previous results that found a LS state for the
metallic \pr\ \cite{Mizokawa1995}. Despite of this strongly
hybridized ground state, the expected value for $S_z$ is close to
the \n\ LS ionic value. The reason is essentially due to the
charge transfer configurations. The amount of charge transfer with
spin up ($S_z=1$ from \underline{$B_1A_1\bf L$}) is compensated by
that of spin down ($S_z=0$ from \underline{$B_1B_1\bf L$}). The
ground state transition takes place at $10D_q=2.1eV$, where we
obtain the same contributions for HS and LS states, that is, the
ground state is made of $\sim$25\%\ \underline{$B_1A_1E$} and of
$\sim$25\%\ \underline{$B_1B_1A_1$}. The spin moment assumes an
intermediate value $S_z\sim0.7$. At the insulating state, with
$10D_q=2.0eV$, the ground state is composed of $\sim$46\%\
\underline{$B_1A_1E$} (HS) and a small amount ($\sim$9\%) of
\underline{$B_1B_1A_1$} (LS), the complement coming from charge
transfer configurations. This leads to a net spin moment
$S_z\sim1.0$. The mixed-spin state arises because these compounds
are between the strong and weak field limit and also due to the
strong hybridization that mixes the Ni3d and O2p orbitals. In case
there would be no charge transfer at all, the high-spin to
low-spin transition would take place over a crystal field range of
less than 0.1 eV. So, it is due to the strong covalence that there
is a gradual change from high-spin to low-spin ground state.

For the \underline{$B_1B_1A_1$} (LS) state, the electron coming
from the O2p ligand band occupies the $B_1$ orbital, even though
it has a higher energy than $A_1$, owing to the gain in exchange
energy. So, \underline{$B_1A_1\bf L$} is the most important charge
transfer configuration at the metallic state. On the other hand,
for the \underline{$B_1A_1E$} (HS) state, the ligand electron must
occupy an orbital already filled and there is no gain in exchange
energy. In this case, the ligand electron occupies the $A_1$
orbital, which is lower in energy than $B_1$ and has a transfer
integral twice larger than for the $E$ orbital. Moreover, the
\underline{$B_1B_1A_1$} (LS) state has all its three holes in
orbitals that have strong mixing with the O2p ligand band ($A_1$
and $B_1$ come from the e$_g$ orbital in O$_h$ symmetry), whereas
the \underline{$B_1A_1E$} (HS) state has a more localized hole in
an $E$ orbital that has less efficient overlap with the O2p ligand
band. So, the increase of the HS component in the ground state and
the decrease of the hybridization for smaller 10D$_q$ are related
to each other.

The physical reason for a smaller 10D$_q$ at the insulating state
originates in the coexistence of two nonequivalent Ni sites and
overall increase of the average Ni-O bonding in all \rn\ compounds
\cite{Munoz1992,Alonso1999,Alonso2000}. Nonequivalent Ni sites
coexist even at the metallic state
\cite{PiamontezePhyScri,PiamontezeLetterEXAFS} and are compatible
with a segregation into two phases, where a more localized
electron phase is embedded in a conducting background
\cite{Zhou2000prb}. The $MI$ transition takes place with the
increase of the less hybridized phase related to the other. Alonso
{\it et al.} \cite{Alonso1999} found for the heavier \rn\
compounds, by refining neutron diffraction data in the monoclinic
symmetry, an antiferromagnetic structure compatible with two
nonequivalent moments. The smaller and larger Ni sites display
magnetic moments of 0.7 and 1.4$\mu_B$, respectively. The values
that we predict here for $S_z$ are larger because the spin moment
of the 3d elements retain most of their isolated ion properties in
x-ray absorption spectroscopy. Owing to the short time scale in
XAS essentially all electrons appear localized. Delocalized
electrons should reduce these values in the solid. The same
argument holds for the angular moment $L_z$, which is normally
quenched in these compounds. Even if somewhat overestimated, our
predictions show that an important parameter controlling the
properties of these compounds is the small splitting in distances,
which leads to a different crystal field in each site. For the
lighter \rn\ powder compounds, the long range structure has always
been refined using a single Ni site and for that reason no such
different spin values have been found. Recently, a monoclinic
distortion was observed in thin films of \nd\ \cite{Staub2002},
questioning the orthorhombic symmetry obtained previously by
neutrons. Our results show that the lighter and heavier \rn\
compounds are very similar from the point of view of the local
electronic structure. Experimental spectra look very similar and
depend only if the compound is insulator or metal. So, we believe
that the same physics observed in the heavier \rn\ compounds may
be involved in the lighter ones.

In conclusion, we presented experimental evidences of significant
changes in the local electronic structure around \n\ in \rn\
compounds. With the support of charge transfer multiplet
calculations, we showed that, concomitantly with the metal to
insulator transition, a decrease of the crystal field splitting
leads to a ground state transition from an essentially LS to a
mixed-spin ground state. The mixed-spin ground state is possible
because the energy separation between HS and LS states is of the
same order of the spin-orbit coupling. The smaller hybridization,
intrinsic to the HS state, gives an explanation to the more
localized character at the insulating regime. The existence of a
mixed-spin state, involving both LS and HS states and also charge
transfer configurations, shed new light in the understanding of
the unusual antiferromagnetic order observed below $T_N$, which is
still an open question.

Work partially supported by LNLS/ABTLuS. CP thanks FAPESP for the
PhD grant. AYR thanks CNPq for the visiting scientist grant.


\end{document}